\documentclass[pra, showpacs, twocolumn]{revtex4}
\usepackage{amssymb}
\usepackage{bm}
\usepackage{graphicx}%

\begin{document}
\title
{Continuous measurements in a composite quantum system and
possible exchange of information between its parts.}

\author{E.\ D.\ Vol}

\affiliation{%
B.Verkin Institute for Low Temperature Physics and Engineering
National Academy of Sciences of Ukraine, Lenin av. 47 Kharkov
61103, Ukraine}

\begin{abstract}
We study an influence of the continuous measurement in a composite
quantum system C on the evolution of the states of its parts. It
is shown that the character of the evolution (decoherence or
recoherence) depends on the type of the measured quantity and on
the initial state of the system. A number of conditions under
which the states of the subsystems of C decohere during the
measuring process are established. We propose a model of the
composite system and specify the observable the measurement of
which may result in the recoherence of the state of one of the
subsystems of C. In the framework of this model we find the
optimal regime for the exchange of information between the parts
of C during the measurement. The main characteristics of such a
process are computed. We propose a scheme of  detection of the
recoherence under the measurement in a concrete physical
experiment.

\end{abstract}

% \submitto{\jpb}
\maketitle

The main goal of the paper is to point out a possibility of the
recoherence of the state of one of the parts of a composite
quantum system under the measurement of physical quantities of a
certain type. Another words, the decoherence of the state of the
composite system, caused by the interaction with the environment,
can be accompanied, under certain conditions, with the
recoherence, or purification, of a mixed state of one of its
parts.

The paper is organized as follows. First, we  review shortly some
background knowledge of the theory of open quantum systems (OQS)
and of the theory of continuous quantum measurements (CQM) used in
the paper. The main object under study, the composite system C
that consists of non-interacting parts, is introduced, and the
evolution of the states of the subsystems of C during the
continuous measurement is considered. A number of conditions under
which the measurement of C results in the decoherence of states of
all its parts are formulated. Then, we consider a more interesting
situation, when  the measurement in the composite system may
result in the recoherence of the state of one  of the subsystems.
In the case when such a recoherence takes place, one can say about
the exchange of information between the subsystems during the
measuring process. On the example of a simple model of the
composite system we study the influence of the initial states of
the parts of C on the process of the exchange of information
between the parts under the measurement of the observable that has
only two eigenvalues and commutes with the Hamiltonian of C. We
find the optimal regime of the information exchange  and compute
the main characteristics of such a process: $\Delta I_{max}$,
$\Delta S $ and $\eta$, where $\Delta I_{max}$ is the maximum
amount of information, received by one subsystem (receiver),
$\Delta S$, the increment of the entropy of the other subsystem
(sender) in the same process, and $\eta=\Delta I_{max}/\Delta S$,
the efficiency coefficient for the information exchange under the
measurement. We also consider the special regime of the
information exchange under which the energies of the subsystems do
not change under the measurement. In the concluding part of the
paper we propose the scheme for the observation of the recoherence
in a concrete physical experiment.

Let us go to the details.

During last years  the point of view, formulated most precisely by
Zurek \cite{1}, becomes more and more widespread in the quantum
physics community. According to it, the behavior of a quantum
system becomes more classical due to its interaction with the
macroscopic environment. As a rule, this interaction results in
the decoherence of the state of the quantum system, i.e., the
transformation of initially pure states into mixed ones. Thus, the
decoherence suppresses the possibility of interference of quantum
states and changes the initial picture of propagating probability
waves to the usual statistical description. The idea of the
decoherence is basic  also for  the theory of continuous quantum
measurements.  The important conceptual advantage of this theory
is the possibility  to consider the measurement as the specific
process of interaction  of two systems:  the measured (quantum)
system and  the measuring (device) system. The measuring system
can be as  classical ones as well as mesoscopic ones, and, in the
framework of the CQM theory,  the description of the measurement
is based on the general principles of quantum mechanics and does
not involve the ideas of the wave function collapse under the
measurement, etc. In the most general form the CQM theory has been
formulated in the papers by Mensky \cite{2} basing on the method
of restricted Feynman path integrals (the method of quantum
corridors). But for the important class of the problems, connected
with the analysis of the behavior of quantum systems under
non-selective measurements, another more simple method can be
used. The method is based on the Lindblad equation and yields the
same results. Let us remind that at non-selective measurements, in
difference with selective ones, we are interesting not in the
result of the measurement, but only in the influence of the
measuring procedure on the state of the measured system. In
general case, the Lindblad equation \cite{3}, that describes the
evolution of the density matrix of an open quantum system (OQS),
has the following form:
\begin{equation}\label{1}
  \frac{d\hat{\rho}}{dt}=-\frac{i}{\hbar}[\hat{H},\hat{\rho}]
  +\frac{\gamma}{2}\left\{[\hat{R}\hat{\rho},\hat{R}^+]
  +[\hat{R},\hat{\rho}\hat{R}^+]\right\}
\end{equation}
(in what follows we  use the systems of units, where $\hbar=1$).

The first term in the right hand side of Eq.(\ref{1}) is connected
with the internal (Hamilton) dynamics of OQS, and the second term
describes its evolution caused by the interaction with the
environment.  A concrete form of the operators $\hat{R}$ and
$\hat{R}^+$ in Eq. (\ref{1}) is determined by reduction of
description of the closed system ("the system under investigation"
+ "the environment") with respect to the environment variables, or
directly from the physical reasons. As was shown, for the first
time, by Lindblad \cite{3}, Eq. (1) for $\hat\rho(t)$ is the only
possible equation of the Markov type that satisfies the main
postulates of quantum mechanics: 1) the superposition principle
(linearity in $\hat{\rho}$); 2) the conservation of the total
probability ($tr (\hat\rho)=1$); and 3) the condition of
non-negativity of $\hat{\rho}$: $\langle \psi
|\hat{\rho}(t)|\psi\rangle \geq 0$ for all $t\geq 0$. Let us show
with the use of Eq. (\ref{1}) how the interaction of the quantum
system with the environment (or the form of the operators
$\hat{R}$ and $\hat{R}^+$) determines the character of evolution
of its states, in particular, their possible decoherence or
recoherence. To do this, we introduce the linear entropy
$S[\hat{\rho}(t)]=tr(\hat{\rho}(t)-\hat{\rho}^2(t))\equiv 1- tr
(\hat{\rho}^2(t))$, the quantity that characterizes a degree of
"purity" (i.e. the coherence) of the quantum state with the
density matrix $\hat{\rho}(t)$. One can show that the linear
entropy has the properties similar to ones of the von Neumann
entropy $S_{vN}=-tr(\hat{\rho}(t) \ln \hat{\rho}(t))$\cite{4}.
But, for the analysis of the evolution of $\hat{\rho}(t)$ in the
framework of Eq. (\ref{1}) the linear entropy serves as more
convenient measure of the coherence of the state. Let us consider
now the important special case of Eq. (\ref{1}), when the
operators $\hat{R}$ and $\hat{R}^+$  commute with each other. One
can prove (see Appendix A) that the linear entropy of such OQS
increases with time $dS[\hat{\rho}]/dt\geq 0$. This result is
valid for any hermitian operator ($\hat{R}=\hat{R}^+=\hat{O}$). In
this case the Lindblad equation (without the Hamilton term) can be
written in the form:
\begin{equation}\label{2}
  \frac{d
  \hat{\rho}}{dt}=-\frac{\gamma}{2}[\hat{O},[\hat{O},\hat{\rho}]].
\end{equation}
According to the CQM theory (see \cite{2}), it is just the
equation that describes the evolution of a quantum system under a
continuous non-selective measurement of an observable $\hat{O}$.
It is necessary to emphasize the difference of  the von Neumann
and the CQM theory schemes of description of the measurement. In
the von Neumann approach the measurement of the observable
$\hat{O}$ is an instant process that results in a sharp change
(collapse) of the state of the system. In the CQM theory the same
final state is reached for the finite time $\Delta t_0\sim
\gamma^{-1}$. This time is  neglected in the von Neumann scheme.
The law of increasing of entropy  ($d S[\hat{\rho}(t)]/dt\geq 0$)
reflects the irreversible character of the measurement  and
witnesses for the decoherence of the state of the measured object.
One can note that, according to Eq. (\ref{1}),  for a certain type
of interaction of OQS with the environment the recoherence of its
state may also take place. As the simplest example, we consider a
two dimensional OQS with the $|g\rangle$ and $|e\rangle$
orthonormal basis states. We choose the operator $|g\rangle\langle
e|$ as the $\hat{R}$ operator, and, correspondingly, the
$|e\rangle\langle g|$ operator as the $\hat{R}^+$ operator. Using
Eq. (\ref{1}) without the Hamilton term, one can show that the
unique stationary state of the system is the pure state
$\hat{\rho}_{st}=|g\rangle\langle g|$. This is the attractive
state:
$\lim_{t\to\infty}\hat{\rho}(t)=\hat{\rho}_{st}=|g\rangle\langle
g|$ for any initial state $\hat{\rho}(0)$. Therefore, beginning
from a certain time, the recoherence of such OQS will take place.
This example can be generalized for OQS of an arbitrary finite
dimension $N$. Moreover, for any fixed pure state of that system
one can construct the corresponding interaction with the
environment that provides evolution of any initial state to that
pure state.

Let us now formulate the problem of interest. We consider a
composite quantum system C consisting of such two parts R and S
that the interaction between these parts can be neglected. Let one
measures an observable $\hat{O}_C$ of the composite system.
According to the foregoing statement,  any state of C  decoheres
during the measurement. But the following question emerges: how
does the states of R and S evolve under the measurement?
Obviously, there are two possibilities (alternatives) A1 and A2 in
such a situation.

A1. The decoherence of the state of the composite system  C is
accompanied with the decoherence of both its parts R and S.

A2. In spite of the decoherence of the  state of C,  the
purification (or the recoherence) of an initial state of one part
(for definiteness, R) of the composite system takes place.

In this paper we are mainly interested in the alternative A2. If
this possibility is realized one can say that an information is
transferred from S to R during the measurement. It is natural to
call the subsystem R as the receiver and the subsystem S as the
sender. Prior the study of the possibility A2 we should point out
a number of conditions that exclude its realization.

One can prove (see Appendix B) that the measurement of an additive
observable $\hat{O}_C=\hat{A}_R\otimes \hat{1}_S+\hat{1}_R\otimes
\hat{B}_S$ results in the decoherence of the state of R and S for
any initial state $\hat{\rho}_C(0)$. Another case of realization
of the possibility A1 is the measurement of a multiplicative
observable $\hat{O}_C=\hat{A}_R\otimes\hat{B}_S$ under condition
that the initial states of the subsystems R and S are
uncorrelated:
$\hat{\rho}_C(0)=\hat{\rho}_R(0)\otimes\hat{\rho}_S(0)$ (see
Appendix B).

Now we consider the simplest model of the composite system C and
construct the observable $\hat{O}_C$ the measurement of which,
under certain initial conditions, realizes the case A2. Let the
Hamiltonian of C (with noninteracting parts R and S) has the form
\begin{equation}\label{3}
  \hat{H}_C=\hat{H}_R\otimes \hat{1}_S + \hat{1}_R \otimes \hat {H}_S,
\end{equation}
where $\hat{H}_R$ and $\hat{H}_S$ are the Hamiltonians of the
subsystems R and S.

Let us assume now that the subsystems S and R are of the same
dimension: $\dim R=\dim S =N$. We also imply that the Hamiltonians
$\hat{H}_S$  and $\hat{H}_R$ are unitary equivalent:
$\hat{H}_S=\hat{U}\hat{H}_R \hat{U}^+$, where
$\hat{U}\hat{U}^+=\hat{U}^+ \hat{U}=\hat{1}$. We define the
operator $\hat{O}_C$ as
\begin{equation}\label{4}
  \hat{O}_C=(\hat{U}^+\otimes \hat{U})\hat{T},
\end{equation}
where $\hat{T}$ is the hermitian operator that permutes the states
of the subsystems R and S. The action of the operator $\hat{T}$ on
the basis states of C is defined by the equation:
\begin{equation}\label{5}
  \hat{T}|i\rangle_R\otimes |j\rangle_S=|j\rangle_R \otimes
  |i\rangle_S.
\end{equation}
Three essential properties of the operator $\hat{O}_C$ follows
directly from the definitions (\ref{4}) and (\ref{5}): a)
$\hat{O}_C=\hat{O}_C^+=\hat{T}(\hat{U}\otimes \hat{U}^+)$; b)
$\hat{O}_C^2=\hat{1}_C$; c) $[\hat{O}_C, \hat{H}_C]=0$.

The property a) shows that $\hat{O}_C$ is the hermitian operator
and, consequently, $\hat{O}_C$ is the observable. It follows from
b) that the eigenvalues of $\hat{O}_C$ are equal to 1 or -1. The
property c) indicates that the total energy of the composite
system C conserves under the measurement of $\hat{O}_C$ (while the
energies of its parts may not conserve). Besides that, we note
that the commutativity of $\hat{O}_C$ and $\hat{H}_C$ results in
that the evolution of the density matrix $\hat{\rho}_C$ is totally
determines by the process of the measurement of $\hat{O}_C$.
Indeed, let us introduce the density matrix $\hat{W}_C$ connected
with $\hat{\rho}_C$ by the unitary transformation
$\hat{\rho}_C=e^{-i\hat{H}_C t}\hat{W}_C e^{i\hat{H}_C t}$. As is
easily seen, the matrix $\hat{W}_C$ satisfies the same equation as
the matrix $\hat{\rho}_C$ but without the Hamilton term:
\begin{equation}\label{6}
  \frac{d \hat{W}_C}{d
  t}=-\frac{1}{2}\left[\hat{O}_C,\left[\hat{O}_C,\hat{W}_C\right]\right].
\end{equation}
Eq. (\ref{6}) for $\hat{W}_C$ is the master equation of the CQM
theory, written in the dimensionless form. Let us introduce now
the density matrices $\hat{\rho}_R$ and $\hat{\rho}_S$ that
describe the states of the parts R and S of the system C and
derive the equations of their evolution under the measurement of
$\hat{O}_C$. By definition, $\hat{\rho}_R=tr_S (\hat{W}_C)$ and
$\hat{\rho}_S=tr_R (\hat{W}_C)$. Taking the traces of the left
hand side and the right hand side of Eq. (\ref{6}) over the states
of the subsystem S and using the properties of the operator
$\hat{O}_C$ we find the equation of  evolution for $\hat{\rho}_R$:
\begin{equation}\label{7a}
  \frac{d \hat{\rho}_R}{d t}= \hat{U}^+\hat{\rho}_S \hat{U} -\hat{\rho}_R.
\end{equation}
Analogously, we obtain the equation for $\hat{\rho}_S$:
\begin{equation}\label{7b}
  \frac{d \hat{\rho}_S}{d t}= \hat{U}\hat{\rho}_R \hat{U}^+ -\hat{\rho}_S.
\end{equation}
Thus, in this model, the measurement of the observable $\hat{O}_C$
(\ref{4}) results in the simple picture of evolution of the states
of the subsystems R and S of the composite system C.  If the
initial states $\hat{\rho}_S(0)$ and $\hat{\rho}_R(0)$ are
specified, the system of equations (\ref{7a}), (\ref{7b}) for
$\hat{\rho}_S(t)$ and $\hat{\rho}_R(t)$ allows to determine the
states of the parts S and R at an arbitrary time $t$. For further
consideration we need to know the relation between the final
states $\hat{\rho}_R(\infty)=\lim_{t\to\infty} {\hat{\rho}}_R(t)$,
 $\hat{\rho}_S(\infty)=\lim_{t\to\infty} \hat{\rho}_S(t)$ and the
 initial states $\hat{\rho}_R(0)$,  $\hat{\rho}_S(0)$.
Integrating the system (\ref{7a}), (\ref{7b}) and approaching the
limit as $t\to \infty$ we obtain the desired relation in the form
of two equations
\begin{equation}\label{8a}
\hat{\rho}_R(\infty)=\frac{\hat{\rho}_R(0)+\hat{U}^+\hat{\rho}_S(0)\hat{U}}{2},
\end{equation}
\begin{equation}\label{8b}
\hat{\rho}_S(\infty)=\frac{\hat{\rho}_S(0)+\hat{U}\hat{\rho}_R(0)\hat{U}^+}{2}.
\end{equation}
Now we have all that is needed for the study of the process of the
exchange of information between the parts R and S under the
measurement of the observable $\hat{O}_C$. We consider this
problem in the following formulation. Let at time $t=0$ (the
starting time of the measurement) the state of the sender
$\hat{\rho}_S(0)$ is known. We will find the initial state of the
receiver $\tilde{\hat{\rho}}_R(0)$ that provides the maximum
amount of information $\Delta I_R$ transferred from S to R during
the measurement. The increment of the amount of information in the
subsystem R for the time of the measurement can be written in the
form:
\begin{eqnarray}\label{9}
  \Delta I_R=-\Delta S_R= S_R \left[ \hat{\rho}_R(0)\right]-S_R
  \left[ \hat{\rho}_R(\infty)\right]\cr = tr (\hat{\rho}_R^2(\infty))- tr
  (\hat{\rho}_R^2(0)).
\end{eqnarray}
Substituting Eq. (\ref{8a}) into Eq. (\ref{9}) we find the
dependence of $\Delta I_R$ on the initial states of R and S
\begin{equation}\label{10}
\Delta I_R=-\frac{3}{4}
tr(\hat{\rho}^2_R(0))+\frac{1}{4}tr(\hat{\rho}^2_S(0))+\frac{1}{2}
tr(\hat{\rho}_R(0)\hat{U}^+\hat{\rho}_S(0) \hat{U}).
\end{equation}
Computing the maximum of the quadratic functional Eq. (\ref{10})
with respect to $\hat{\rho}_R(0)$ under the additional restriction
$tr(\hat{\rho}_R(0))=1$ we determine the desired density matrix
$\tilde{\hat{\rho}}_R(0)$
\begin{equation}\label{11}
  \tilde{\hat{\rho}}_R(0)=\frac{1}{3} \hat{U}^+ \hat{\rho}_S(0)
  \hat{U}+\frac{2}{3N}\hat{1}_R
\end{equation}
and the maximum amount of information $\Delta I_R$ transferred
from S to R in the optimal regime for the given initial state of
the part S
\begin{equation}\label{12}
  \Delta
  I^{max}_R\left\{\hat{\rho}_S(0)\right\}=\frac{1}{3}
  \left[tr(\hat{\rho}^2_S(0))-\frac{1}{N}\right].
\end{equation}
As follows from the expression (\ref{12}), for any initial state
of the sender (excluding the disordered one
$\hat{\rho}_S(0)=\hat{1}_S/N$) one can find initial states
$\hat{\rho}_R(0)$ for which the amount of information $\Delta I$
received during the measurement will be positive. The global
maximum of the amount of information transferred from S to R is
reached for the pure initial state of S and it is equal to
\begin{equation}\label{13}
  \Delta I_{max}=\frac{1}{3}\left(1-\frac{1}{N}\right)
\end{equation}
Another essential characteristics of the information transfer
process is the increment of the entropy of the sender $\Delta S$.
Let us find this quantity for the optimal regime of the
information exchange considered above.
\begin{eqnarray}\label{14}
  \Delta S \equiv S \left[ \hat{\rho}_S(\infty)\right]-S\left[
  \hat{\rho}_S(0)\right] \cr= tr(\hat{\rho}_S^2(0))-
  tr(\hat{\rho}_S^2(\infty))\cr =\frac{3}{4}
  tr(\hat{\rho}_S^2(0))-\frac{1}{4}
  tr(\tilde{\hat{\rho}}_R^2(0))
  -\frac{1}{2} tr(\hat{\rho}_S(0)\hat{U} {\tilde{\hat{\rho}}}_R(0)\hat{U}^+)
\end{eqnarray}
Note that we use Eq. (\ref{8b}) for the derivation of (\ref{14}).
Substituting the expression (\ref{11}) for
$\tilde{\hat{\rho}}_R(0)$ into Eq. (\ref{14}) we obtain
\begin{equation}\label{15}
  \Delta S \left[
  \hat{\rho}_S(0)\right]=\frac{5}{9}\left[tr(\hat{\rho}_S^2(0))-\frac{1}{N}\right].
\end{equation}
Following the approach accepted in the thermodynamics of the
informational processes (see \cite{6}) we introduce the
coefficient of efficiency of the information transfer $\eta$. By
definition, $\eta\equiv \Delta I_R/\Delta S$.  It follows from the
law of the increase of entropy  that $\eta\leq 1$ for any
evolution of the composite system. Comparing the expressions
(\ref{12}) and (\ref{15}) we find that  $\eta_0=3/5$ and it does
not depend on the initial state of the sender in the optimal
regime of the information exchange.

Let us now discuss the question on the relation between the energy
transfer and the information transfer under the measurement of
$\hat{O}_C$ in the model considered. For this purpose we determine
the evolution of the average energies of the subsystems R and S:
$E_R(t)\equiv tr(\hat{\rho}_R(t)\hat{H}_R)$ and $E_S(t)\equiv
tr(\hat{\rho}_S(t)\hat{H}_S)$. Using Eqs. (\ref{7a}), (\ref{7b}),
we obtain the simple equations of evolution of $E_R(t)$ and
$E_S(t)$:
\begin{equation}\label{16a}
  \frac{d E_R}{d t}= E_S- E_R,
\end{equation}
\begin{equation}\label{16b}
  \frac{d E_S}{d t}= E_R- E_S.
\end{equation}
Note that the unitary equivalence of the Hamiltonians $H_R$ and
$H_S$ was used for the derivation of (\ref{16a}), (\ref{16b}). It
follows from these equations that the energy of the composite
system conserves and the energy is transferred from the "hotter"
to "colder" part until their energies becomes equal each other at
the end of the measurement. It is clear that if the initial
energies of the parts are equal each other, there is no transfer
of energy under the measurement. We call such a regime the
isoenergetic one. Let us compute the main characteristics of the
information transfer process in the optimal isoenergetic regime.
As the matter of fact, in this case, the formulation of the
problem and the method of computing of the quantities $\Delta
I_R$, $\Delta S$ and $\eta$ remains the same. For instance, to
compute $\Delta I_R^e$ one should find the maximum of the
functional (\ref{10}) with respect to $\hat{\rho}_R(0)$ under two
additional conditions: $tr(\hat{\rho}_R(0))=1$ and
$tr(\hat{\rho}_R(0) \hat{H}_R)=tr(\hat{\rho}_S(0) \hat{H}_S)$.
Using the Lagrange multipliers method we obtain,  after simple
computations, the optimal initial state of the part R
\begin{equation}\label{17}
  \hat{\rho}_R^e(0)=\frac{\hat{U}^+ \hat{\rho}_S(0)\hat{U}}{3}+\frac{2}{3
  N}\hat{1}_R+\frac{2}{3}\frac{ tr(\hat{\rho}_S(0)\hat{H}_S)}{tr(\hat{H}_R^2)}
  \hat{H}_R.
\end{equation}
Substituting this expression  into equation (\ref{10})  we obtain
the maximum amount of information transferred in the isoenergetic
regime of the measurement of $\hat{O}_C$
\begin{equation}\label{18}
  \Delta I_R^e \left\{
  \hat{\rho}_S(0)\right\}=\frac{1}{3}\left\{tr(\hat{\rho}_S^2(0))-\frac{1}{N}-
  \frac{\left(tr(\hat{\rho}_S(0)
  \hat{H}_S)\right)^2}{tr(\hat{H}_R^2)}\right\}
\end{equation}
Note that the derivation of (\ref{18}) was done under the
additional condition $tr(\hat{H}_R)=0$ that fixes the reference
point for the energy. This condition does not influence on the
generality of the results obtained.

We also present the result for the increment of the entropy of the
sender S in the isoenergetic regime of the measurement of
$\hat{O}_C$:
\begin{equation}\label{19}
  \Delta S^e \left\{ \hat{\rho}_S(0)\right\}=\frac{5}{9} \left[
  tr(\hat{\rho}_S^2(0))-\frac{1}{N}-\frac{\left(tr(\hat{\rho}_S(0)
  \hat{H}_S)\right)^2}{tr(\hat{H}_R^2)}\right]
\end{equation}
Comparing the expressions (\ref{18}) and (\ref{19}), we find that
the coefficient of efficiency of the information transfer in such
a regime $\eta$ is also equal to 3/5.

To illustrate the expression (\ref{18}) we consider the optimal
process of the measurement for the simplest composite system C of
the dimension 4 ($\dim C =4$, $\dim R=\dim S=2$) without energy
exchange between its parts. We use the representation where the
Hamiltonian of the sender is diagonal: $
\hat{H}_S=\Delta_0\left(\matrix{1&0\cr 0&-1}\right)$  with
$2\Delta_0$, the distance between the energy levels of the part S.
Let at the starting time of the measurement  the state of S is
described by the density matrix
$\hat{\rho}_S(0)=\left(\matrix{a&c\cr c^*&b}\right)$. Then
$tr(\hat{\rho}_S \hat{H}_S)=\Delta_0(a-b)$,
$tr(\hat{H}_R^2)=tr(\hat{H}_S^2)= 2 \Delta_0^2$ and under
accounting the normalization condition $a+b=1$ the expression
(\ref{18}) for $\Delta I^e_R$ takes the form
\begin{eqnarray}\label{20}
  \Delta
  I^e_R\left\{\hat{\rho}_S(0)\right\}=\frac{1}{3}\left[a^2+b^2+2|c|^2-\frac{1}{2}
  -\frac{(a-b)^2}{2}\right]\cr =\frac{2}{3}|c|^2.
\end{eqnarray}
One can see that the maximum amount of information transferred to
one of the parts under the measurement of $\hat{O}_C$ in the
isoenergetic regime is determined by the non-diagonal elements of
the density matrix $\hat{\rho}_S(0)$ in the representation where
$\hat{H}_S$ is diagonal.

Thus, we summarize the results of the paper.  On the example of
the simple model of the measurement in the composite system we
have demonstrated the possibility of the exchange of information
between its parts. We compute the main characteristics of such a
process for the optimal initial states of the subsystems. One
should emphasize that, in itself, the possibility of the
recoherence of one part of the composite system under the
measurement is determined entirely by the type of the measured
quantity and by the initial state of the system, and it does not
depend on the simplified assumptions used in our consideration.
The generalization of the results to the case of unitary
non-equivalent Hamiltonians of the parts R and S and the study of
the information exchange process between the subsystems of
different dimensions is postponed for further publications. In
conclusion, we discuss shortly the possibility of the observation
of the effect predicted in this paper - the recoherence of the
state of the subsystem under the continuous measurement. For the
first time, the suggestion to use the procedure of continuous
measurement of the energy of a two-level system for the monitoring
of the quantum transition was put forward in Ref. \cite{7}. The
experimental scheme of realization of this idea was also described
in that paper. The object of the measurement is a polarized atom
with the transition excited by resonant pumping. The electron beam
that scatters on the atom (interacting with its dipole momentum)
is used as a meter. Measuring the scattering angle one can obtain
the energy of the probed atom at any time.

Using the ideas of Ref. \cite{7}, we describe the simplest, from
our point of view, scheme of the experiment in which the effect of
recoherence under the measurement in a composite system can
emerge. For this purpose, we consider two crossed beams of neutral
particles (neutrons) with the spins $s_1=s_2=1/2$ that propagate
close to each other in the area that  contains a massive magnetic
atom with the spin $S$. We imply the neutrons of the beam 1 are in
the mixed state described by the density matrix $\hat{\rho}_1$ and
the neutrons of the beam 2 - in the state described by the density
matrix $\hat{\rho}_2$. It is assumed that two beams are
synchronized in such a way that in time when the neutron of one
beam moves close to the magnetic atom there is also the neutron of
beam 2 near this atom. Since there is the exchange interaction
between neutrons of different beams and between neutrons and the
magnetic atom, one can consider that in such an experiment the
massive atom provides the continuous measurement of the observable
$\hat{O}=\hat{{\bf s}}_1 \hat{{\bf s}}_2$ of the two-particle
system. Let us remind now that the operator $\hat{{\bf s}}_1
\hat{{\bf s}}_2$ is connected with the operator of spin
permutation $\hat{T}$ by the relation $\hat{T}=(1+\hat{{\bf s}}_1
\hat{{\bf s}}_2/4)/2$ (see \cite{8}). Taking into account the
foregoing statements of the paper we arrive to the conclusion that
under proper choice of the states of the beams $\hat{\rho}_1$ and
$\hat{\rho}_2$ such an experiment should demonstrate the effect of
recoherence of the state of neutrons of one of the beams.
Comparing the interference pattern of the recoherred beam with the
interference pattern of the test beam (obtained for the same
$\hat{\rho}_1$ and $\hat{\rho}_2$, but without the "measuring"
magnetic atom) one can check all main conclusions and relations of
this paper.

I would like to acknowledge I.V.Krive, L.A.Pastur and A.A.Zvyagin
for useful discussions of the results of the paper.

\appendix
\section{}
Let us present the proof of the statement of the paper on the
monotonical increase of (linear) entropy of OQS,  the interaction
of which with the environment is described in the framework of the
Lindblad  equation (\ref{1}) by the operators $\hat{R}$ and
$\hat{R}^+$ that obey the relation $[\hat{R},\hat{R}^+]=0$.

Using Eq. (\ref{1}) and the expression for the linear entropy
$S[\hat{\rho}]=1-tr(\hat{\rho}^2)$ one can write the expression
for the rate of change of the entropy
\begin{equation}\label{a1}
  \frac{d S}{d t}= -2 tr(\hat{\rho}\dot{\hat{\rho}})=
  2\gamma tr(\hat{R}^+\hat{R}
  \hat{\rho}^2-\hat{\rho}\hat{R}\hat{\rho}\hat{R}^+).
\end{equation}
As follows from (\ref{a1}) the monotonic increase of the entropy
($dS/dt\geq 0$) takes place if the following inequality is
satisfied:
\begin{equation}\label{a2}
  tr(\hat{R}^+\hat{R}\hat{\rho}^2)\geq
  tr(\hat{\rho}\hat{R}\hat{\rho}\hat{R}^+).
\end{equation}
To prove the inequality (\ref{a2}) we use the
Cauchy-Bunyakovsky-Schwarz (CBS) inequality
\begin{equation}\label{a3}
  ||{\bf A}||^2\cdot||{\bf B}||^2\geq ({\bf A B})^2,
\end{equation}
that is valid in any linear space where the scalar product of two
vectors ${\bf A}$ and ${\bf B}$ ($||{\bf A}||\equiv\sqrt{{\bf A
A}}$) is defined. We remind that for any pair of linear operators
$\hat{A}$ and $\hat{B}$ acting in a vector space of finite
dimension one can define the operation $\hat{A}\hat{B}\equiv
tr(\hat{A}\hat{B}^+)$ that satisfies all axioms of the scalar
product \cite{9}. We choose the operator $\hat{\rho}\hat{R}$ as
$\hat{A}$ and the operator $\hat{R}\hat{\rho}$ as $\hat{B}$ and
write the CBS inequality (\ref{a3}) for these operators:
\begin{equation}\label{a4}
  tr(\hat{R}\hat{R}^+\rho^2)tr(\hat{R}^+\hat{R} \hat{\rho}^2)\geq
  [tr(\hat{\rho}\hat{R}\hat{\rho}\hat{R}^+)]^2.
\end{equation}
Here we use the possibility to do cyclic permutations of operators
under the trace.

Using the commutativity of $\hat{R}$ and $\hat{R}^+$ and taking
into account that
$tr(\hat{R}^+\hat{R}\hat{\rho}^2)=tr(\hat{\rho}\hat{R}\hat{R}^+\hat{\rho})\geq
0$ we obtain from (\ref{a4}) the required inequality (\ref{a2}).
Thus, the statement on the monotonic increase of linear entropy in
such OQS is proven.

\section{}
 In this appendix we consider two particular realizations (cases) of the alternative
 A1 (the decoherence of the states of the parts under the
 continuous measurement  in the composite system).

 Case 1. The measurement of the additive variable.

Under such  measurement the evolution  of the state
$\hat{\rho}_C(t)$ is described by the master equation of the CQM
theory. This equation, written in the dimensionless form, reads as
\begin{equation}\label{b1}
  \frac{ d \hat{\rho}_C(t)}{d
  t}=-\frac{1}{2}\left[\hat{O}_C,\left[\hat{O}_C,\hat{\rho}_C\right]\right],
  \end{equation}
where
\begin{equation}\label{b2}
  \hat{O}_C=\hat{A}_R\otimes \hat{1}_S + \hat{1}_R\otimes\hat{B}_S.
\end{equation}
One can check directly that the general solution of Eq. (\ref{b1})
has the form
\begin{equation}\label{b3}
  \hat{\rho}_C(t)=\frac{1}{\sqrt{2\pi t}}\int_{-\infty}^{\infty} d
  s e^{-s^2/2t} e^{-i \hat{O}_C s}\hat{\rho}_C(0) e^{i \hat{O}_C
  s}.
\end{equation}
The factor $\exp(-i \hat{O}_C s)$ under the integral in (\ref{b3})
can be presented as
\begin{eqnarray}\label{b4}
\exp(-i \hat{O}_C s)=\exp(-i(\hat{A}_R\otimes \hat{1}_S +
\hat{1}_R\otimes \hat{B}_S)s)\cr =(\exp(-i \hat{A}_R s)\otimes
\hat{1}_S)\cdot(\hat{1}_R\otimes\exp(-i \hat{B}_S s)).
\end{eqnarray}
According to the quantum theory, $\hat{\rho}_R(t)$, the state of
the part $R$, is determined by the relation
\begin{equation}\label{b5}
  {\rho}_{ij}^R(t)=tr_S(\hat{\rho}_C(t))\equiv\sum_\alpha
  \langle i\alpha|\hat{\rho}_C(t)|j\alpha\rangle,
\end{equation}
where $|i\alpha\rangle\equiv|i\rangle_R\otimes|\alpha\rangle_S$ is
the orthogonal basis in C.

Using the expressions (\ref{b3}), (\ref{b4}) and the definition
(\ref{b5}) we find
\begin{eqnarray}\label{b6}
  {\rho}_{ij}^R(t)=\frac{1}{\sqrt{2\pi t}}\int_{-\infty}^{\infty} d
  s e^{-\frac{s^2}{2t}}\sum_\alpha
  \langle i\alpha| e^{-i \hat{O}_C s}\hat{\rho}_C(0) e^{i \hat{O}_C
  s}|j\alpha\rangle\cr=\sum_{kl}\frac{1}
  {\sqrt{2\pi t}}\int_{-\infty}^{\infty} d
  s e^{-s^2/2t} \langle i|e^{-i \hat{A}_R s}|k\rangle\cr \cdot \sum_\gamma
  \langle k \gamma|\hat{\rho}_C(0)| l \gamma \rangle\langle l| e^{i \hat{A}_R
  s}|j\rangle\cr =\frac{1}{\sqrt{2\pi t}}\int_{-\infty}^{\infty} d
  s e^{-s^2/2t} \left(e^{-i \hat{A}_R s}\hat{\rho}_R(0) e^{i \hat{A}_R
  s}\right)_{ij}.
\end{eqnarray}
Thus, one can see that the measurement of the additive quantity
$\hat{O}_C$ is, in fact, reduced to the measurement of the
quantities $\hat{A}_R$ and $\hat{B}_S$ in each of the subsystems
and that is why it results in the decoherence of its states.

Case 2. The measurement of the multiplicative observable.

As above, the evolution of $\hat{\rho}_C(t)$ is determined by Eq.
(\ref{b1}), but the measured quantity $\hat{O}_C$ has the form
\begin{equation}\label{b7}
  \hat{O}_C=\hat{A}_R\otimes \hat{B}_S.
\end{equation}
One can use the general form (\ref{b3}) of the solution of Eq.
(\ref{b1}) and write $\hat{\rho}_C(t)$ in the basis $|i \rangle_R
\otimes |\alpha\rangle_S$, where $|i\rangle_R$ are the
eigenvectors of $\hat{A}_R$, and  $|\alpha\rangle_S$ are the
eigenvectors of $\hat{B}_S$. After simple calculations we find
that
\begin{equation}\label{b8}
 \langle i\alpha|\hat{\rho}_C(t)|j\beta\rangle =
  \langle i\alpha|\hat{\rho}_C(0)|j\beta\rangle e^{-(A_i
  B_\alpha-A_j B_\beta)^2 t},
\end{equation}
where $A_i$ and $B_\alpha$ are the eigenvalues of the operators
$\hat{A}_R$ and $\hat{B}_S$ in the states $|i\rangle_R$ and
$|\alpha\rangle_S$, respectively.

Let us compute the rate of change of  the entropy of the subsystem
R
\begin{eqnarray}\label{b9}
  \frac{d S_R}{d t} =-\frac{d}{d t} tr(\hat{\rho}_R^2(t))=-2 \sum_{ij}
  \rho_{ij}^R(t)\dot{\rho}_{ji}^R(t)\cr=2\sum_{\alpha\beta i j} e^{-(B_\alpha^2+B_\beta^2)
  (A_i-A_j)^2 t} (A_i-A_j)^2 B_\beta^2 \cr \times \langle i\alpha|\hat{\rho}_C(0)
  |j\alpha\rangle\langle j\beta|\hat{\rho}_C(0) |i\beta\rangle
\end{eqnarray}
If one assumes that at the starting time of the measurement of
$\hat{O}_C$ the state of the parts R and S are not correlated
($\hat{\rho}_C(0)=\hat{\rho}_R(0)\otimes\hat{\rho}_S(0)$), the
expression (\ref{b9}) is reduced to
\begin{eqnarray}\label{b10}
 \frac{d S_R}{d t}= 2\sum_{\alpha\beta i j} (A_i-A_j)^2 B^2_\beta  e^{-(B_\alpha^2+B_\beta^2)
  (A_i-A_j)^2 t}\cr \times |{\rho}_{ij}^R(0)|^2
  {\rho}_{\alpha\alpha}^S(0){\rho}^S_{\beta\beta}(0).
\end{eqnarray}
One can see directly from (\ref{b10}) that $d S_R/ dt\geq 0$. The
relation $d S_S/ dt\geq 0$ can be obtained by the same way. Thus,
we have proven the statement of the paper on the decoherence of
the state of the subsystem R and S under the measurement of the
multiplicative observable $\hat{O}_C$ for the case of uncorrelated
initial states of R and S.

%\section*{References}

\end{document}